\documentclass[12pt]{article}

\usepackage[left=3cm,right=3cm,top=2cm,bottom=2cm]{geometry} 
\usepackage[font={small}, width=14cm]{caption}
\usepackage{amsmath}
\usepackage{amssymb}
\usepackage{amsthm}
\usepackage{amscd}
\usepackage{bbm}
\usepackage{enumitem}
\usepackage{float}
\usepackage{authblk}
\usepackage{xcolor}
\usepackage{mathrsfs}
\usepackage[mathscr]{eucal}
\usepackage[normalem]{ulem}
\usepackage{graphicx}
\usepackage{tikz-cd}
\usepackage{comment}
\usepackage{pgfplots}
\usepackage{subcaption}
 \usepackage{slashed}
 
\usetikzlibrary{cd}

\usepackage{pgfplots}

\pgfplotsset{compat=1.10}
\usepgfplotslibrary{fillbetween}
\usetikzlibrary{patterns}

\newtheorem{thm}{Theorem}[section]
\newtheorem{prop}[thm]{Proposition}

\newtheorem{cor}[thm]{Corollary}

\theoremstyle{definition}

\theoremstyle{remark}
\newtheorem{rem}[thm]{Remark}

\newtheorem{example}[thm]{Example}

\newcommand{\CC}{\mathbb{C}}
\newcommand{\NN}{\mathbb{N}}
\newcommand{\RR}{\mathbb{R}}

\newcommand{\TT}{\mathbb{T}}

\newcommand{\Tr}{\mathrm{Tr}}

\title{Fractional index of Bargmann--Fock space and Landau levels}

\author{Guo Chuan Thiang
}
\affil{\normalsize Beijing International Center for Mathematical Research, Peking University, China \\ ORCiD: 0000-0003-0268-0065}

\begin{document}

\maketitle
\abstract{The lowest Landau level Hilbert space, or the Bargmann--Fock space, admits a quantized trace for the commutator of its position coordinate operators. We exploit the Carey--Pincus theory of principal functions of trace class commutators to probe this integer quantization result further, and uncover a hidden rational structure in the higher-order commutator-traces. This shows how exact fractional quantization can occur whenever exact integral quantization does.}

\bigskip

\section{Overview}
The (Bargmann--Segal--)\emph{Fock space} $\mathscr{H}$ is the space of holomorphic functions on the complex plane $\CC\cong\RR^2$, which are square-integrable with respect to Gaussian-weighted Lebesgue measure $e^{-|z|^2}d\mu$. It has been studied extensively since its introduction in \cite{Bargmann} over sixty years ago. As recalled in Section \ref{sec:LL}, $\mathscr{H}$ arises in quantum physics as the lowest eigenspace of the Landau Hamiltonian (magnetic Laplacian), introduced over ninety years ago. The Landau Hamiltonian is the simplest model of quantum dynamics of non-interacting electrons in a plane subject to a uniform perpendicular magnetic field. Over forty years ago, it was discovered experimentally \cite{vonKlitzingQHE} that such 2D electron systems exhibit \emph{exact integer quantization} of their Hall conductance in units of $e^2/h$, where $e$ is the electron charge. It is generally accepted that electrons filling up the states in $\mathscr{H}$ contribute \emph{exactly} one unit of Hall conductance. A brief discussion of existing rigorous calculations verifying this statement is provided in Remark \ref{rem:folklore}. Further fractional conductances were subsequently discovered \cite{Tsui}, but these have not been understood or calculated at the same level of rigour as the integer case.

In this paper, we report on the remarkable index-theoretic features of $\mathscr{H}$, guided by the exact integer and fractional quantizations discovered in physics experiments mentioned above. As a warm up, we provide a direct calculation of the Hall conductance of $\mathscr{H}$ by elementary means. This quantized calculation is then placed in the context of the general Carey--Helton--Howe--Pincus (CHHP) theory of traces of commutators. The CHHP theory was influential in the development of noncommutative geometry and cyclic cohomology. The importance of the latter (and the integer Fredholm index) for the integer quantum Hall effect is well-known \cite{BES, ASS}. To illustrate the power of the CHHP theory, we prove that higher-order analogues of the trace formula for Hall conductance are necessarily \emph{rational}. Unlike the integer case, these rational trace results do not seem to be computable or deducible without appealing to the CHHP theory. Finally, we speculate on how this ``hidden'' rational trace structure might improve our understanding of exact fractional quantization of Hall conductance, particularly its modelling by anyonic quasiparticles.

\section{Trace of commutator of Toeplitz operators on Fock space}
\subsection{Bargmann--Fock space}\label{sec:BFspace}
The Bargmann--Fock space $\mathscr{H}$ is a Hilbert space with inner product
\[
(\varphi, \psi)=\frac{1}{\pi}\int_\CC\overline{\varphi(z)}\psi(z)e^{-|z|^2}\,d\mu(z).
\]
It will be convenient to attach Gaussian weights $e^{-|z|^2/2}$ to the Fock space functions, thus $\mathscr{H}$ is regarded as a Hilbert subspace of $L^2(\CC)\equiv L^2(\CC,\mu)$ with the standard Lebesgue measure $\mu$. So, for example, the functions
\begin{equation}
z\mapsto z^ne^{-|z|^2/2},\qquad n=0,1,\ldots\label{eqn:LLL.eigenfunctions}
\end{equation}
form an orthogonal basis for $\mathscr{H}\subset L^2(\CC)$. We write
\[
P:L^2(\CC,\mu)\to\mathscr{H}
\]
for the orthogonal projection. Its integral kernel is
\begin{align}
P(z,w)&=\frac{1}{\pi}e^{-\frac{|z|^2+|w|^2}{2}}e^{z\bar{w}}\label{eqn:integral.kernel}\\
&=\frac{1}{\pi}e^{-\frac{1}{2}|z-w|^2}e^{iw\wedge z},\qquad w\wedge z:=w_1z_2-z_1w_2,\label{eqn:integral.kernel.form2}
\end{align}
which is smooth and exponentially decaying away from the diagonal.

\subsection{Fock--Toeplitz operators and their commutator-traces}

On the full Hilbert space $L^2(\CC)$, each $f\in L^\infty(\CC)$ acts as a bounded operator by pointwise multiplication (which we denote with the same symbol $f$). These $f$ are bounded functions of the standard $x_1$ and $x_2$ position coordinate operators, and they obviously commute with each other. However, the compressions of $f$ to $\mathscr{H}={\rm Range}(P)$, i.e., the \emph{Fock--Toeplitz operators}
\[
P_f:=PfP,
\]
no longer commute in general.

Intuitively, non-commutativity of the Fock--Toeplitz operators $P_f$ arises because they act on a holomorphic subspace $\mathscr{H}\subset L^2(\CC)$ having a preference for the counter-clockwise orientation on $\CC$. In other words, the range of $P$, i.e., $\mathscr{H}$, has \emph{chiral asymmetry}, and we may quantify this asymmetry by the following commutator-trace. Let $f_1, f_2$ denote the characteristic functions supported on the right half-plane and the upper half-plane, respectively. The difference between the ``counter-clockwise product'' $P_{f_2}P_{f_1}$ and the ``clockwise product'' $P_{f_1}P_{f_2}$ is numerically quantified by
\begin{equation}
\Tr\big[P_{f_1},P_{f_2}\big]=\Tr\big(P_{f_1}P_{f_2}-P_{f_2}P_{f_1}\big)=\Tr\big(Pf_1Pf_2P-Pf_2Pf_1P\big).\label{eqn:chiral.asymmetry}
\end{equation}
As recalled in Proposition \ref{prop:main.calculation}, the commutator $[P_{f_1},P_{f_2}]$ is indeed trace class. Higher-order versions of \eqref{eqn:chiral.asymmetry} will be investigated in Section \ref{sec:rational}.

\begin{rem}\label{rem:Lidskii}
The only way for $\Tr[P_{f_1},P_{f_2}]\neq 0$ to occur, is if $P_{f_1}P_{f_2}$ and $P_{f_2}P_{f_1}$ are \emph{not} individually trace class. For otherwise, they must have the same trace, and $\Tr[P_{f_1},P_{f_2}]$ would vanish. Therefore, whenever $\Tr[P_{f_1},P_{f_2}]\neq 0$, a na\"{i}ve computation
\[
``\Tr[P_{f_1},P_{f_2}]=\Tr(P_{f_1}P_{f_2})-\Tr(P_{f_2}P_{f_1})"
\]
will fail.
\end{rem}

\begin{figure}[h]
\centering
\begin{tikzpicture}[yscale=0.5]
\begin{axis}[axis x line=middle,
      		axis y line=left,
            xlabel=$x$,
            xtick=\empty,
            ytick={0,1}]
\addplot[name path=F, ultra thick, dashed] table {
0  0
1  0
3  1
5  1
}node[pos=.4, above left]{$\Lambda(x+a)$};
\addplot[name path=G, ultra thick] table {
0  0
2  0
4  1
5  1
}node[pos=.5, above right]{$\quad\Lambda(x)$};

\addplot[pattern=horizontal lines]fill between[of=F and G, soft clip={domain=1:4}]
;
\node[coordinate,pin=30:{$A$}] at (axis cs:3.8,3){};

\end{axis}
\end{tikzpicture}\caption{For a switch function $\Lambda$, the integral \eqref{eqn:switch.integral} is the signed area between the graph of $\Lambda$ and the graph of its $a$-translate, which equals $a$.}\label{fig:switch.integral}
\end{figure}
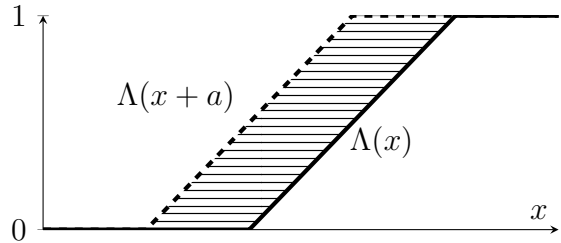

More generally, let us define a \emph{switch function} to be a non-decreasing measurable function $\Lambda:\RR\to[0,1]$, such that there exists an interval $[c,d]$ with
\begin{equation}
\Lambda(x)=\begin{cases} 0, & x<c,\\ 1, & x>d.\end{cases}\label{eqn:switch.interval}
\end{equation}
Switch functions $\Lambda$ generalize the Heaviside step function. As observed in Prop.\ 6.9 of \cite{ASS}, a basic property of switch functions is
\begin{equation}
\int_{x\in\RR} \Lambda(x+a)-\Lambda(x)=a,\qquad \forall \,a\in\RR,\label{eqn:switch.integral}
\end{equation}
see Fig.\ \ref{fig:switch.integral}.

The quantization result \eqref{eqn:basic.commutator} below, is well-known in the context of the quantized Hall conductance. For example, a closely related calculation appears in \cite{ASS} in connection with a Fredholm index; see Remark \ref{rem:folklore} for a more detailed discussion and comparison with \eqref{eqn:basic.commutator}. We find it conceptually and pedagogically clearer to compute \eqref{eqn:basic.commutator} by direct elementary methods, \emph{before} providing the index-theoretic meaning of the quantized result in Section \ref{sec:index.theoretic}.

\begin{prop}\label{prop:main.calculation}
Let $\Lambda_1,\Lambda_2:\RR\to[0,1]$ be any switch functions, so
\[
f_1(u):=\Lambda_1(u_1),\qquad f_2(u):=\Lambda_2(u_2),\qquad\quad (u=u_1+iu_2\in\CC)
\]
are switch functions of the first and second coordinates, respectively. Then 
\begin{equation}
\Tr[P_{f_1},P_{f_2}]=\frac{1}{2\pi i}.\label{eqn:basic.commutator}
\end{equation}
\end{prop}

\begin{proof} 
\begin{enumerate}[label={{Step} \arabic*.}, leftmargin=*]
    \item We first rewrite
\begin{align}
[P_{f_1},P_{f_2}]&=Pf_1Pf_2P -Pf_2Pf_1P\nonumber\\
&=P(f_1P-Pf_1)(f_2P-Pf_2)-P(f_2P-Pf_2)(f_1P-Pf_1)\nonumber\\
&=P[f_1,P][f_2,P]-P[f_2,P][f_1,P],\label{eqn:rewritten}
\end{align}
so that \emph{products} of commutators appear in each term on the right. For $i=1,2$, write $[c_i,d_i]\subset\RR$ for the interpolation region of $\Lambda_i$, as in \eqref{eqn:switch.interval}. Notice that
   \[
   [f_1,P]=f_1P(1-f_1)-(1-f_1)Pf_1,
   \]
with $\mathrm{Supp}(f_1)\subset[c_1,\infty)\times\RR$ and $\mathrm{Supp}(1-f_1)\subset(-\infty,d_1]\times\RR$. This means that the operator $[f_1,P]$ is supported near the vertical strip $[c_1,d_1]\times\RR$ in $\CC$. More precisely, in view of Eq.\ \eqref{eqn:integral.kernel.form2}, the integral kernel of $[f_1,P]$ decays exponentially away from this strip. Similarly, $[f_2,P]$ is supported near the horizontal strip $\RR\times [c_2,d_2]$. So $[f_1,P][f_2,P]$ is supported near the compact square $[c_1,d_1]\times [c_2,d_2]$, see Fig.\ \ref{fig:localization}. Consequently, $[f_1,P][f_2,P]$ is trace class; see Prop.\ 3.7 of \cite{TX} for a direct verification, and \cite{LTpartition} for a coarse geometry account of such arguments. This means that the trace of \eqref{eqn:rewritten} makes sense, and may be evaluated by the integral
\begin{align}
\Tr[P_{f_1},P_{f_2}]&=   \int_{u\in\CC}\big(P[f_1,P][f_2,P]\big)(u,u)-(1\leftrightarrow 2)\nonumber\\
&=\int_{u,v,w\in\CC} P(u,v)P(v,w)P(w,u)\big(f_1(v)-f_1(w)\big)\big(f_2(w)-f_2(u)\big)\nonumber\\
&\qquad\qquad\qquad\qquad\qquad\qquad\qquad\qquad\;-(1\leftrightarrow 2).\label{eqn:kernel.modified}
\end{align}

\begin{figure}[h]
\centering
\begin{tikzpicture}
\fill[lightgray] (-0.8,-2) rectangle (0.8,2);
\fill[pattern=north east lines] (-3,-0.5) rectangle (3,0.5);
\node at (0,-1.5) {$\mathrm{Supp}(f_1^\prime)$};
\node at (2,0) {$\mathrm{Supp}(f_2^\prime)$};
\end{tikzpicture}\caption{Let $f_1, f_2$ be smooth switch functions in the first and second coordinates of $\CC$, respectively. The integral kernel of $[f_1,P][f_2,P]$ is concentrated near the intersection of the interpolation regions for $f_1$ and $f_2$.}\label{fig:localization}
\end{figure}
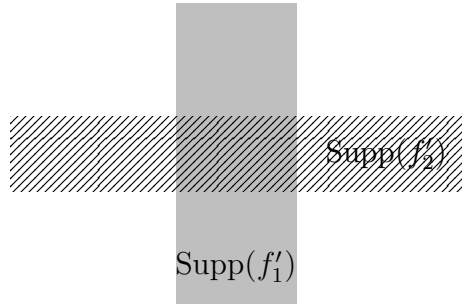

    \item Formula \eqref{eqn:integral.kernel.form2} shows that the integral kernel of $P$ enjoys the translation covariance
\begin{equation*}
P(u-t,v-t)=U_t(u)P(u,v)U_t(v)^{-1},\qquad \forall\, t\in \CC,
\end{equation*}
where $U_t(u)=e^{iu\wedge t}$. In particular, setting $t=w$, we deduce that
\begin{align}
P(u,v)P(v,w)P(w,u)&=P(u-w,v-w)P(v-w,0)P(0,u-w)\nonumber\\
&=P(u-w,v-w)\cdot\frac{1}{\pi^2}e^{-\frac{|u-w|^2+|v-w|^2}{2}}.\label{eqn:triple.invariance}
\end{align}
Substituting \eqref{eqn:triple.invariance} into \eqref{eqn:kernel.modified} and changing variables from $u,v$ to $u+w,v+w$, we obtain
\begin{align}
{\rm Tr}[P_{f_1},P_{f_2}]&=\frac{1}{\pi^2}\int_{u,v,w\in\CC}P(u,v)e^{-\frac{|u|^2+|v|^2}{2}}\big(f_1(v+w)-f_1(w)\big)\big(f_2(w)-f_2(u+w)\big)\nonumber \\
&\qquad\qquad\qquad\qquad\qquad\qquad\qquad\qquad\qquad\qquad\qquad\qquad\qquad -(1\leftrightarrow 2).\nonumber
\end{align}

\item The $w$-integral above is easily carried out with the help of \eqref{eqn:switch.integral},
 \begin{align}
 &\int_{w\in\CC}\big(f_1(v+w)-f_1(w)\big)\big(f_2(w)-f_2(u+w)\big)\;\;-(1\leftrightarrow 2)\nonumber\\
 \quad&=\int_{w_1\in\RR}\big(\Lambda_1(v_1+w_1)-\Lambda_1(w_1)\big)\int_{w_2\in\RR}\big(\Lambda_2(w_2)-\Lambda_2(u_2+w_2)\big)\;\;-(1\leftrightarrow 2)\nonumber\\
 \quad&=v_1(-u_2)-v_2(-u_1)\equiv u\wedge v.\nonumber
 \end{align}
A double integral remains,
 \begin{align}
 \Tr[P_{f_1},P_{f_2}]&=\frac{1}{\pi^2}\int_{u,v\in\CC}P(u,v)e^{-\frac{|u|^2+|v|^2}{2}}u\wedge v\nonumber\\
 &\overset{\eqref{eqn:integral.kernel}}{=}\frac{1}{\pi^3}\int_{u,v\in\CC}e^{-(|u|^2+|v|^2)}e^{u\bar{v}}u\wedge v,\label{eqn:last.integral.formula}
 \end{align}
and it can be evaluated by passing to polar coordinates \mbox{$u=r_1e^{i\theta_1}, v=r_2e^{i\theta_2}$},
 \begin{align}
&{\rm Tr}[P_{f_1},P_{f_2}]\nonumber\\
&=\frac{1}{\pi^3}\int_{r_1,r_2\in[0,\infty)}r_1r_2\,e^{-(r_1^2+r_2^2)}\int_{\theta_1,\theta_2\in[0,2\pi]}e^{r_1r_2e^{i(\theta_1-\theta_2)}}r_1r_2\sin(\theta_2-\theta_1)\nonumber\\
&=\frac{1}{\pi^3}\int_0^\infty\int_0^\infty dr_1\,dr_2\,(r_1r_2)^2e^{-(r_1^2+r_2^2)}\int_{0}^{2\pi}d\theta_1\int_{S^1}\frac{d{z}}{i{z}}\,e^{r_1r_2\bar{{z}}}\cdot\frac{{z}-\bar{{z}}}{2i}\qquad\;({z}:=e^{i(\theta_2-\theta_1)})\nonumber\\
&=\frac{1}{\pi^3}\int_0^\infty\int_0^\infty dr_1\,dr_2\,(r_1r_2)^2e^{-(r_1^2+r_2^2)}\cdot(2\pi)\cdot(2\pi i)\big(\frac{r_1r_2}{-2}\big)\quad\qquad\qquad\quad\;({\rm residue\; theorem})\nonumber\\
&=\frac{2}{\pi i}\Bigg(\underbrace{\int_0^\infty dr\,r^3e^{-r^2}}_{1/2}\Bigg)^2=\frac{1}{2\pi i}.\nonumber
\end{align}
\end{enumerate}
\end{proof}

\section{Index-theoretic connections and rational traces}\label{sec:index.theoretic}
Proposition \ref{prop:main.calculation} should be understood as the basic input for a general index-theoretic analysis, in the following sense.

\subsection{Carey--Helton--Howe--Pincus index theory}
A pair of bounded self-adjoint operators $A,B$ is said to \emph{almost commute}, if $[A,B]$ is trace class. Historically, examples with $\Tr[A,B]\neq 0$ came from singular integral operators on $\RR$, and Toeplitz operators on the circle or on (higher-dimensional) Bergman spaces of $L^2$-analytic functions on the disc \cite{HH,HHacta,TWZ}. Prop.\ \ref{prop:main.calculation} indicates that Toeplitz operators on the Bargmann--Fock space provide new examples with direct physical meaning, and we will soon see that this class of examples has a remarkable commutator-trace structure.

\medskip
Let $p,q$ be complex-coefficient polynomials in two real variables $x,y$. Since $[A,B]$ is trace class, the operators $[p(A,B),q(A,B)]$ are also trace class. Although there is an ordering ambiguity in the operators $p(A,B)$ and $q(A,B)$, a little algebra reveals that $\Tr[p(A,B),q(A,B)]$ is unambiguous due to cyclicity of the trace. For instance,
\[
[AB,B]-[BA,B]=[[A,B],B]=\underbrace{[A,B]B}_{\text{trace class}}-\underbrace{B[A,B]}_{\text{trace class}}=\text{traceless}.
\]
So the assignment
\[
(p,q)\mapsto \Tr[p(A,B),q(A,B)]
\]
is well-defined, and is antisymmetric bilinear in $p,q$. It is natural to seek a general formula for $\Tr[p(A,B),q(A,B)]$ for \emph{all} polynomials $p,q$, not just for the base case $\Tr[A,B]$. A reasonable guess might be expected to involve the Poisson bracket 
\[
\{p,q\}=\frac{\partial p}{\partial x_1}\frac{\partial q}{\partial x_2}-\frac{\partial p}{\partial x_2}\frac{\partial q}{\partial x_1}.
\]

In the 1970s, Helton--Howe and Carey--Pincus obtained deep results in this direction \cite{HH,CP-PNAS}. Despite an obvious analogy to the idea of quantization of Poisson brackets, their theory does not seem to be known in the physics community, so we take this opportunity to advertise their key results. The following are taken from \cite{CP-PNAS}.
\begin{enumerate}
\item\label{item:1} There exists a compactly supported Borel measure $\mu_{A,B}$ on $\CC$, such that
\begin{equation}
\Tr[p(A,B),q(A,B)]=i\int_{\CC}\{p,q\}\,d\mu_{A,B}.\label{eqn:CP.formula}
\end{equation}
Here, $p,q$ are arbitrary polynomial functions of the real and imaginary parts of \mbox{$x+iy\in\CC$}. Thus, on the right side of \eqref{eqn:CP.formula}, the dependence on $p,q$ is decoupled from the dependence on $A,B$, with the former appearing in the classical Poisson bracket and the latter appearing in the integration measure.
\item The measure $\mu_{A,B}$ is absolutely continuous with respect to Lebesgue measure $\mu$, with Radon--Nikodym derivative the so-called \emph{principal function}
\[
\frac{1}{2\pi}G_{A,B}=\frac{d\mu_{A,B}}{d\mu}.
\]
Furthermore, whenever $z=x+iy\in\CC$ lies outside the essential spectrum of $A+iB$, 
\begin{equation}
G_{A,B}(z)=\mathrm{Index}(A+iB-z).\label{eqn:Pincus.index}
\end{equation}
It will be convenient to write the CHHP trace formula \eqref{eqn:CP.formula} in terms of the principal function,
\begin{equation}
\Tr[p(A,B),q(A,B)]=\frac{i}{2\pi}\int_{\CC}\{p,q\}\,G_{A,B}\,d\mu.\label{eqn:HH.formula}
\end{equation}
\end{enumerate}

A brief account of the history of formulae \eqref{eqn:HH.formula} and \eqref{eqn:Pincus.index} can be found in \cite{HoweBook}. 

\begin{example}\label{ex:Toeplitz}
Consider the unilateral shift $S$ on $\ell^2(\NN)={\overline{\rm span}}\{e_0, e_1, e_2\ldots\}$. The operator $S$ is an example of a Toeplitz operator $T_f$ on the Hardy space $H^2(\TT)$ (the subspace of $L^2(\TT)$ with non-negative Fourier coefficients), where the symbol function $f:\TT\to \CC$ is just the identity function $z\mapsto z$ on the unit circle $\TT$. For general $f\in C^\infty(\TT)$, the Toeplitz operator $T_f$ is defined to be the compression of the multiplication-by-$f$ operator on $L^2(\TT)$ to an operator on $H^2(\TT)$. It is clear that $[S^*,S]$ is the rank-1 projection onto the span of $e_0$. Following \S I.1 of \cite{HH}, it is not hard to build on this and derive the \emph{Helton--Howe formula}
\begin{equation}
{\rm Tr}[T_f,T_g]=\frac{1}{2\pi i}\int_D d\tilde{f}\wedge d\tilde{g}=\frac{1}{2\pi i}\int_D\{\tilde{f},\tilde{g}\}\,dxdy.\label{eqn:HH.Toeplitz}
\end{equation}
Here, $\tilde{f}, \tilde{g}$ are smooth extensions of $f,g$ to the unit disc $D$. 

Write
\begin{equation}
A=\frac{1}{2}(S+S^*),\qquad B=\frac{1}{2i}(S-S^*),\label{eqn:Toeplitz.AC.pair}
\end{equation}
for the real and imaginary parts of $S$. Consider Laurent polynomials $\tilde{f},\tilde{g}$ of the complex variable $z=x+iy$; we may view them as polynomials $p,q$ of two real variables $x,y$. The left side of \eqref{eqn:HH.Toeplitz} can be replaced by ${\rm Tr}[p(A,B), q(A,B)]$, so that
\begin{equation}
{\rm Tr}[p(A,B),q(A,B)]=\frac{1}{2\pi i}\int_D\{p,q\}\,dxdy.\label{eqn:HH.Toeplitz.real}
\end{equation}
Now, the essential spectrum of $A+iB=S$ is the unit circle $\TT$, and continuity of Fredholm index gives
\[
{\rm Index}(A+iB-z)=\begin{cases}{\rm Index}(S)=-1,& |z|<1,\\
0, & |z|>1.
\end{cases}
\]
So the right side of \eqref{eqn:HH.Toeplitz.real} does coincide with the CHHP formulae \eqref{eqn:Pincus.index}--\eqref{eqn:HH.formula} for the almost commuting pair \eqref{eqn:Toeplitz.AC.pair}.
\end{example}

\begin{rem}\label{rem:intuition}
In \cite{HH}, Helton--Howe studied a vast generalization of \eqref{eqn:HH.Toeplitz.real} for general almost commuting pairs $A,B$. Given any such $A,B$, they showed that there exists some measure $d\mu_{A,B}$ on the plane, such that \eqref{eqn:CP.formula} holds. The existence of the representation \eqref{eqn:CP.formula} hinges on the fact that the \emph{bilinear} functional
\[
\mathcal{B}_{A,B}:(p,q)\mapsto {\rm Tr}[p(A,B),q(A,B)]
\]
satisfies the so-called \emph{collapsing property}, meaning that $\mathcal{B}_{A,B}(r_1(s)),r_2(s))=0$ whenever $r_1,r_2$ are single-variable polynomials of the same two-variable polynomial $s$. Then, invoking an algebraic lemma of Wallach (see \S I.2 of \cite{HH} for a proof), there must exist a \emph{linear} functional $\mathcal{L}_{A,B}$ such that
\[
\mathcal{B}_{A,B}(p,q)=\mathcal{L}_{A,B}\{p,q\}
\]
holds, for arbitrary two-variable polynomials $p,q$.

In \cite{HH}, it was also shown that at $z$ not belonging to the essential spectrum of $A+iB$, the representing measure $id\mu_{A,B}$ equals $\frac{i}{2\pi}d\mu$ times the Fredholm index of $A+iB-z$. Generally, $A+iB$ will have essential spectrum. Even if the Lebesgue measure of this essential spectrum is zero (as was the case in Example \ref{ex:Toeplitz}), it is not immediate that its $P$-measure is zero as well. The important technical result on the absolute continuity of $dP$, thus the existence of the principal function representation \eqref{eqn:HH.formula}, was established by Carey--Pincus \cite{CP-PNAS}.
\end{rem}

\subsection{Rational quantization of commutator-trace of Fock--Toeplitz operators}\label{sec:rational.quantization.Fock.Toeplitz}
It is a considerable challenge to find naturally occurring examples of $A,B$ for which $G_{A,B}$ can be fully determined. It is therefore remarkable that we can completely determine $G_{A,B}$ for the almost-commuting pair of Fock--Toeplitz operators
\[
A=P_{f_1},\qquad B=P_{f_2},
\]
using just the single calculation of Prop.\ \ref{prop:main.calculation} as input. Let us explain how this works. (Compare with the case of classical Toeplitz operators in Example \ref{ex:Toeplitz}, and see \cite{TX} for further examples.) 

\medskip

Let $S=[0,1]\times [0,1]\subset \CC$ be the unit square.
The operator $P_{f_1}+iP_{f_2}=P_{f_1+if_2}$ has spectrum $S$, and essential spectrum
\begin{equation*}
\mathrm{ess}\textit{-}\mathrm{spec}(P_{f_1+if_2})=\partial S,\label{eqn:square.ess.spec}
\end{equation*}
see \cite{TX}, Theorem 4.1, for a proof.
Crucially, the essential spectrum $\partial S$ has zero Lebesgue measure. In accordance with the formula \eqref{eqn:Pincus.index}, $G_{P_{f_1},P_{f_2}}$ is thereby completely determined, as a $L^1(\CC)$ class, by the Fredholm indices of $P_{f_1+if_2}-z$ for $z\in \CC\setminus\partial S$. For large $z\in\CC$, the operator $P_{f_1+if_2}-z$ becomes invertible, so its index vanishes. By continuity of Fredholm index, the index of $P_{f_1+if_2}-z$ vanishes for all $z\in \CC\setminus S$. Thus $G_{P_{f_1},P_{f_2}}$ is supported on the unit square. By continuity of Fredholm index again, we have ${\rm Index}(P_{f_1+if_2}-z)$ being the same integer $n$ for all $z$ in $S\setminus \partial S$. Thus we conclude that
\begin{equation*}
G_{P_{f_1},P_{f_2}}=n\cdot\chi_S,\qquad n=\mathrm{Index}\big(P_{f_1+if_2}-z_0\big),\quad z_0\in S\setminus \partial S,\label{eqn:Pincus.function.for.Fock}
\end{equation*}
where $\chi_S$ denotes the characteristic function on the unit square $S$. It remains to determine what the integer $n$ is.

Substituting $G_{P_{f_1},P_{f_2}}=n\cdot\chi_S$ into the base case of the CHHP trace formula \eqref{eqn:HH.formula}, i.e., $p\equiv p(x,y)=x$ and $q\equiv q(x,y)=y$, we obtain the index formula
\begin{equation}
\Tr[P_{f_1},P_{f_2}]=\frac{i}{2\pi }\int_\CC n\cdot\chi_S\, d\mu=\frac{in}{2\pi},\qquad n=\mathrm{Index}\big(P_{f_1+if_2}-z_0\big).\label{eqn:calculation.is.index}
\end{equation}
In Prop.\ \ref{prop:main.calculation}, we computed the left side of \eqref{eqn:calculation.is.index} to be $-\frac{i}{2\pi}$, thus we obtain $n=-1$. We have now fully determined the principal function to be
\begin{equation}
G_{P_{f_1},P_{f_2}}=-\chi_S.\label{eqn:Landau.principal.function}
\end{equation}

Because the principal function has such a simple form, the CHHP trace formula for general $p,q$, is likewise very simple:
\begin{thm}\label{thm:generalization}
Let $P$ be the Fock space projection and $f_1,f_2:\CC\to[0,1]$ be any switch functions in the first and second coordinates respectively. Then for any polynomials $p,q$ in two variables,
\[
\Tr\left[p(P_{f_1},P_{f_2}),q(P_{f_1},P_{f_2})\right]=\frac{1}{2\pi i}\int_S\{p,q\},
\]
where $S$ is the unit square $[0,1]\times[0,1]\subset\CC$ equipped with Lebesgue measure, and $\{\cdot,\cdot\}$ is the Poisson bracket.
\end{thm}

The appearance of the unit square $S$ in Theorem \ref{thm:generalization} has the following important consequence.
\begin{cor}\label{cor:rational}
If in the setup of Theorem \ref{thm:generalization}, the polynomials $p,q$ have rational coefficients, then
\begin{equation}
2\pi i\,\Tr\left[p(P_{f_1},P_{f_2}),q(P_{f_1},P_{f_2})\right]=\int_S\{p,q\}\,\in\,\mathbb{Q}.\label{eqn:rational}
\end{equation}
\end{cor}
\begin{proof}
By the assumption on $p,q$, the Poisson bracket $\{p,q\}$ is a two-variable polynomial with $\mathbb{Q}$-coefficients. So the integral of $\{p,q\}$ over the unit square $S=[0,1]\times[0,1]$ is a rational number. Theorem \ref{thm:generalization} shows that this integral exactly computes the trace on the left side of \eqref{eqn:rational}.
\end{proof}

\begin{rem}\label{rem:general.b}
The Gaussian weight in the definition of Bargmann--Fock space can be modified to $e^{-b|z|^2/4}$, where $b>0$ is a real parameter corresponding to the applied magnetic field strength (see Section \ref{sec:LL}). Proposition \ref{prop:main.calculation}, and the statement that the principal function for $P_{f_1},P_{f_2}$ equals $-\chi_S$, remain true for any $b>0$.
\end{rem}

\section{Exact quantization of Hall conductance}
In this section, we explain the significance of Theorem \ref{thm:generalization} and Corollary \ref{cor:rational} to the quantum Hall effect.

\subsection{Hall conductance of Landau levels}\label{sec:LL}
Let $H=H^*$ be a self-adjoint magnetic Schr\"{o}dinger operator on $L^2(\CC)$. Here, $\CC$ is regarded as the planar sample manifold hosting a 2D gas of non-interacting electrons, and $H$ is the energy operator for a single such electron when subjected to a magnetic field. Because electrons are fermions obeying the Pauli exclusion principle, at zero temperature, they occupy the eigenspaces of $H$ in ascending energy order, until the so-called \emph{Fermi energy} $E$ is attained. The value of $E$ depends on material properties such as the density of electrons, and the density of single-particle states afforded by $H$. The \emph{Fermi projection} is the spectral projection $P^{(E)}=\chi_{(\infty,E]}(H)$, and it describes the quantum mechanical ground state of the non-interacting electron gas at zero temperature. 

Let $f_1, f_2$ be switch functions of the $x$ and $y$ coordinates respectively. Provided $E$ lies in a spectral gap of $H$, the expression
\begin{align}
\sigma_{\rm Hall}\big(P^{(E)}\big)&:=-i\Tr\Big(P^{(E)}\big[[f_1,P^{(E)}],[f_2,P^{(E)}]\big]\Big)\label{eqn:Kubo.linear.response}\\
&\overset{\eqref{eqn:rewritten}}{=}-i\Tr\big[P^{(E)}_{f_1},P^{(E)}_{f_2}\big],& P^{(E)}_{f_i}=P^{(E)}f_iP^{(E)},\label{eqn:Kubo.formula}
\end{align}
is the \emph{Kubo trace formula} for the \emph{Hall conductance} of $P^{(E)}$. See \cite{ES, DEF, ASS} for derivations of the formula \eqref{eqn:Kubo.linear.response} based on linear response theory. We focus on the equivalent commutator trace form, Eq.\ \eqref{eqn:Kubo.formula}, which emphasizes the chiral asymmetry of $P^{(E)}$. Eq.\ \eqref{eqn:Kubo.formula} plays a prominent role in A.\ Kitaev's generalization of ``Chern number'' \cite{Kitaev} for the study of models exhibiting anyonic excitations (see Appendix C--D of \cite{Kitaev}), and is studied systematically in \cite{LTpartition} as a coarse index pairing.

\medskip

The fundamental source of interesting $P^{(E)}$ is the magnetic Laplacian $H_b$, or \emph{Landau operator}, for a 2D electron gas subject to a magnetic field of uniform strength $b>0$ perpendicular to the plane $\CC$. Geometrically, there is a Hermitian line bundle $\mathcal{L}^\nabla\to\CC$ whose connection $\nabla$ has curvature 2-form $-ib\cdot dx\wedge dy=\frac{b}{2}d\bar{z}\wedge dz$. The $\mathfrak{u}(1)=i\RR$-valued connection 1-form can be taken to be $\mathcal{A}=\frac{b}{4}(z\,d\bar{z}-\bar{z}\,dz)=\frac{ib}{2}(y\,dx-x\,dy)$, and the Landau operator is concretely given by the self-adjoint operator
\[
H_b=(d+\mathcal{A})^*(d+\mathcal{A})=-(\partial_x+\tfrac{iby}{2})^2-(\partial_y-\tfrac{ibx}{2})^2.
\]
The spectrum of $H_b$ is readily solved by considering the associated Dirac operator twisted by $\mathcal{L}^\nabla$,
\begin{equation}
D_b=-2i\begin{pmatrix}0 & {\partial}-\frac{b}{4}\bar{z}\\ \bar{\partial} +\frac{b}{4}{z} & 0\end{pmatrix},\qquad \partial=\frac{1}{2}(\partial_x-i\partial_y),\; \bar{\partial}=\frac{1}{2}(\partial_x+i\partial_y).\label{eqn:twisted.Dirac}
\end{equation}
From the Schr\"{o}dinger--Lichnerowicz identity,
\[
D_b^2=\begin{pmatrix}H_b-b & 0 \\ 0 & H_b+b\end{pmatrix},
\]
it follows readily that 
\begin{equation}
\mathrm{Spec}(H_b)=\{(2\ell+1)b\,:\, \ell=0,1,\ldots\}.\label{eqn:Landau.quantization}
\end{equation}
This is the well-known \emph{Landau quantization} \cite{Landau} of the Laplace spectrum into a discrete set when $b>0$ is turned on. Each eigenvalue is called a \emph{Landau level}, and is infinitely-degenerate. For example, the lowest Landau level (LLL) eigenspace is 
\begin{equation}
\ker(H_b-b)=\ker(D_b)=\ker(\bar{\partial}+\tfrac{b}{4}z).\label{eqn:LLL.Dirac.kernel}
\end{equation}
For $b=2$, the LLL eigenspace \eqref{eqn:LLL.Dirac.kernel} is precisely the Fock space $\mathscr{H}$ defined in Section \ref{sec:BFspace}. As explained in Remark \ref{rem:general.b}, the same holds for general $b>0$, with the appropriate modification of the Gaussian exponent in the definition of Fock space. Thus the Fock space projection $P$ is exactly the Fermi projection of $H_b$ when the Fermi energy lies just above the LLL eigenvalue $b$.

\medskip
Now, $P$ describes the state where the electrons fully occupy the LLL eigenspace, i.e., the \emph{filling factor} $\nu$ equals the integer $1$. According to experiments, the Hall conductance of $P$ should equal $-\frac{\nu}{2\pi}=-\frac{1}{2\pi}$, meaning that $n$ should equal $-1$. Indeed, Proposition \ref{prop:main.calculation} is precisely the calculation that
\begin{equation}
\sigma_{\rm Hall}(P)=-i\Tr[P_{f_1},P_{f_2}]=-\frac{1}{2\pi},\label{eqn:folklore}
\end{equation}
verifying that the Landau Hamiltonian model and its Landau levels lead to the correct Hall conductance prediction in the $\nu=1$ case.

\medskip

The result \eqref{eqn:folklore} is known by other methods, but our approach gives further information --- the principal function. In turn, we easily deduce a new result, Corollary \ref{cor:rational}, which indicates that there are hidden \emph{fractional traces} associated to $P_{f_1},P_{f_2}$, if we look carefully enough. We will discuss this in Section \ref{sec:rational}. For now, we briefly remark on some earlier approaches to Eq.\ \eqref{eqn:folklore}.

\begin{rem}\label{rem:folklore}
The authors of \cite{ASS} used their theory of \emph{relative index of projections} to prove their Theorem 6.8, that 
\begin{equation}
2\pi i\,\tilde{\Tr}[P_{f_1},P_{f_2}]=-\mathrm{Index}\Big(P\frac{z}{|z|}P\Big).\label{eqn:charge.deficiency}
\end{equation}
Using the explicit basis \eqref{eqn:LLL.eigenfunctions} for the range of $P$, the Fredholm index on the right side of \eqref{eqn:charge.deficiency} was read off as $-1$. However, the trace class property of $[P_{f_1},P_{f_2}]$ was not established in \cite{ASS}. Instead, a limit of spatially-truncated traces, denoted $\tilde{\Tr}$ above, was used, see pp.\ 414 of \cite{ASS}. The trace class property is not trivial to establish, see Prop.\ 3.7 of \cite{TX} and \cite{ES} for direct analytic proofs, and \cite{LTpartition} for a general coarse-geometric argument.
The first steps in our proof of Proposition \eqref{prop:main.calculation} are a simplified version of the preparatory steps made in \cite{ASS} to establish their Eq.\ \eqref{eqn:charge.deficiency}.

The integral \eqref{eqn:last.integral.formula} also appears in \cite{BES} Lemma 5, which states that the so-called \emph{Chern character} ${\rm Ch}(P)$ equals $-1$. The Chern character was identified with a related Kubo formula applicable to \emph{homogeneous} projections, and required use of a \emph{trace-per-unit-area} instead of the ordinary trace. We follow \cite{DEF,ES} and use the ordinary trace, noting that it is physically important and mathematically possible to handle non-homogeneous projections \cite{LTpartition}.
\end{rem}

\subsubsection{Stability and macroscopic nature of index}
Eq.~\eqref{eqn:charge.deficiency} from \cite{ASS} is historically significant as it connected the experimental \emph{stability} of the integer-quantized values of $\sigma_{\rm Hall}$ to the stability of a Fredholm index. Our formula \eqref{eqn:calculation.is.index} gives another identification of Hall conductance with a different Fredholm index. In this latter regard, a vast generalization was recently achieved in \cite{LTpartition}: the expression $2\pi i\Tr[P_{f_1},P_{f_2}]$ was proved to be a certain \emph{integral} pairing in coarse index theory. Briefly, the pairing depends only on the coarse (co)homology of the data $P, f_1,f_2$, and is therefore extremely stable, even against small-scale changes of metric and/or magnetic field strength on the planar manifold. 

Furthermore, as discussed in \cite{LTpartition}, spatially localized projections always give rise to trivial coarse index pairings. So the non-triviality of $2\pi i\Tr[P_{f_1},P_{f_2}]$ reveals a \emph{macroscopic} index-theoretic property of $P$ (manifested as its quantized Hall conductance), which is stable against \emph{microscopic} perturbations.

\subsubsection{Magnetic lattice fermions}\label{sec:lattice.fermions}
We have focused on \emph{Landau levels} of magnetic Schr\"{o}dinger operators, particularly those of the magnetic Laplacian. The well-definedness of the Kubo trace formula \eqref{eqn:Kubo.formula} can also be established for Fermi projections $P=P_{\rm lattice}$ of certain \emph{short-range, lattice model} Hamiltonians for 2D electrons in a magnetic field, as long as the Fermi energy $E$ lies in a spectral gap (or more generally, mobility gap) of the lattice Hamiltonian $H$; see, e.g., \cite{EGS}. In such models, the integral quantization of the Kubo trace formula is well-established; see, e.g., Appendix of \cite{EGS} and \cite{BES}. There are also recent arguments that lattice models of interacting electrons can exhibit fractionally quantized conductances \cite{Bachmann}.

Given that $[P_{f_1},P_{f_2}]$ is also trace class in many lattice models, CHHP theory tells us that in such a setting, there would again a principal function $G_{P_{f_1},P_{f_2}}$ such the trace formula \eqref{eqn:HH.formula} holds. The question is whether we also have a simple expression
\begin{equation}
G_{P_{f_1},P_{f_2}}\overset{?}{=}n\cdot\chi_S,\label{eqn:lattice.principal.function}
\end{equation}
as was the case for $P=P_{\rm LLL}$ the LLL projection (Eq.~\eqref{eqn:Landau.principal.function}). If \eqref{eqn:lattice.principal.function} holds, then the rational trace result, Corollary \ref{cor:rational}, will follow, and the possible relation to fractionally quantized conductances discussed in Section \ref{sec:rational} applies.

By the arguments of Section \ref{sec:rational.quantization.Fock.Toeplitz}, to establish Eq.~\eqref{eqn:lattice.principal.function}, it suffices to show that
\[
\mathrm{ess}\textit{-}\mathrm{spec}(P_{f_1+if_2})\overset{?}{\subseteq}\partial S.
\]
Since $P_{f_1}$ and $P_{f_2}$ have spectrum in $[0,1]$ and commute up to trace class (thus compact) operators, we certainly have $\mathrm{ess}\textit{-}\mathrm{spec}(P_{f_1+if_2})\subseteq S$. So the task is reduced to checking whether
\begin{equation}
P_{f_1+if_2}-z\quad\text{is invertible modulo compacts}\;\;\;\forall\, z\in S\setminus \partial S.\label{eqn:Fredholm.inverse.existence}
\end{equation}
For the case of $P=P_{\rm LLL}$, the property \eqref{eqn:Fredholm.inverse.existence} was proved in Theorem 4.1 of \cite{TX}. The proof there relied on establishing estimates guaranteeing that $P=P_{\rm LLL}$ is locally trace class, and has sufficiently well-controlled propagation; this class of projections was studied in detail in \cite{LTpartition}. For Fermi projections coming from lattice model Hamiltonians, depending on the precise model assumptions, one could mimic the argument in \cite{TX} to try to establish \eqref{eqn:Fredholm.inverse.existence}, then \eqref{eqn:lattice.principal.function} and rational trace results would follow. We leave a systematic analysis of lattice models to future work.

\subsection{Higher Landau levels and additivity}\label{sec:additivity}
Let $P_j$ denote the $j$-th Landau level eigenprojection. Theorem \ref{thm:generalization} remains true with $P=P_0$ replaced by $P_j$ \cite{TX}. This is again well-known to physicists, and perhaps obvious from the ladder operator method used to deduce Landau quantization. Corollary \ref{cor:rational} readily follows for each $P_j$ as well. 

It is not so obvious that the Kubo formula for $\sigma_{\rm Hall}(\cdot)$ is additive, i.e.,
\begin{equation}
\Tr\left[(P\oplus P^\prime) f_1(P\oplus P^\prime),(P\oplus P^\prime) f_2 (P\oplus P^\prime)\right]=\Tr[P_{f_1},P_{f_2}]+\Tr[P^\prime_{f_1}, P^\prime_{f_2}],\label{eqn:basic.additivity}
\end{equation}
since multiple factors of $P,P^\prime$ appear in the commutator on the left side; see \cite{TX} for a proof. Even in the translation-invariant situation where the Kubo formula can be understood as a Chern number of a vector bundle of occupied states, additivity of Chern numbers is a standard but non-obvious differential geometric fact. Additivity, together with the above-mentioned fact that each Landau level contributes equally to the Hall conductance, is needed for consistency with the experimental result that the Hall conductance at integral filling factor $\nu=\ell\in\NN$ equals $\ell$ times that at $\nu=1$.

For each non-negative integer $\ell\in\NN$, let
\[
P^{(\ell)}:=\bigoplus_{j=0}^\ell P_j.
\]
Thus $P^{(\ell)}$ is the Fermi projection when the Fermi energy lies between the $\ell$-th and $(\ell+1)$-th Landau level. Write
\[
P_{j,f}:=P_jfP_j,\qquad P^{(\ell)}_{f}:=P^{(\ell)}fP^{(\ell)}
\]
for the compressions of $f$ to the ranges of $P_j$ and $P^{(\ell)}$. Let $f_1, f_2$ be switch functions of the first and second coordinates, respectively, as in Proposition \ref{prop:main.calculation}.

In \cite{TX}, we proved that $[P^{(\ell)}_{f_1},P^{(\ell)}_{f_2}]$ is trace class, with trace $\frac{\ell+1}{2\pi i}$, and that $P^{(\ell)}_{f_1+if_2}$ still has essential spectrum being $\partial S$. The same argument as in the $\ell=0$ case gives $-(\ell+1)\cdot\chi_S$ as the principal function for the pair $P^{(\ell)}_{f_1},P^{(\ell)}_{f_2}$. Then the analogue of Theorem \ref{thm:generalization} for $P^{(\ell)}$ is
\begin{align*}
\Tr\left[p\big(P^{(\ell)}_{f_1},P^{(\ell)}_{f_2}\big),q\big(P^{(\ell)}_{f_1},P^{(\ell)}_{f_2}\big)\right]&=\frac{\ell+1}{2\pi i}\int_S\{p,q\}\\
&=\sum_{j=0}^\ell\Tr\left[p\big(P_{j,f_1}, P_{j,f_2}\big),q\big(P_{j,f_1}, P_{j,f_2}\big)\right].
\end{align*}
In other words, for arbitrary polynomials $p,q$, the assignment
\[
P_j\mapsto \Tr[p(P_{j,f_1},P_{j,f_2}),q(P_{j,f_1},P_{j,f_2})]
\]
continues to be additive in the Landau level projections $P_j$. This general additivity property is a new result, and does not seem to be deducible without appealing to the principal functions.

\subsection{Exact numerical quantization}\label{sec:exact.quantization}
Dirac's canonical quantization scheme is summarized by the canonical commutation relation (CCR)
\begin{equation}
[\hat{X},\hat{P}]=i\hbar\cdot\mathbf{1}.\label{eqn:CCR}
\end{equation}
Here, $\hat{X},\hat{P}$ are unbounded self-adjoint operators ``quantizing'' the position and momentum coordinate functions $x,p$ on classical phase space, and $\mathbf{1}$ is the identity operator on the representation Hilbert space $L^2(\RR)$. Formally, Eq.~\eqref{eqn:CCR} is obtained by replacing $x,p$ with $\hat{X},\hat{P}$, and replacing the Poisson bracket of classical phase space functions with the commutator of Hilbert space operators. One quickly encounters several issues when dealing with \eqref{eqn:CCR}. First, the meaning of the commutator of unbounded operators can be ambiguous, see \S VIII.5 of \cite{RS1}. Second, there are unavoidable ordering issues which make the correspondence of $[\cdot,\cdot]$ with Poisson brackets ambiguous at higher orders in $\hbar$. For example, $xpx=x^2p=px^2$ but $\hat{X}\hat{P}\hat{X}\neq \hat{X}^2\hat{P}\neq \hat{P}\hat{X}^2$. The ambiguities become more serious for higher-order polynomials.

Ordering issues arise for general CCRs, not just the standard example of \eqref{eqn:CCR}. For example, in the quantum Hall effect setting, one often encounters arguments saying that when the (multiplication operators by the) position coordinate functions $x, y$ are projected onto a Landau level, they become noncommuting ``position coordinate operators'' $\hat{x}, \hat{y}$ obeying a CCR,
\begin{equation}
[\hat{x},\hat{y}]=i\hbar/eb,\label{eqn:formal.QHE.NCG}
\end{equation}
see \cite{SW} and \S V.A of \cite{DN}. Here, $b$ is the magnetic field strength, and we included the reduced Planck's constant $\hbar$ and the electron charge $e$. Another pair of ``noncommuting coordinates'' is the ``\emph{guiding centre coordinates}''
\[
R_1=\frac{1}{2}x-\frac{i\hbar}{eb}\partial_y,\qquad R_2=\frac{1}{2}y+\frac{i\hbar}{eb}\partial_x,
\]
also obeying a CCR,
\begin{equation}
[R_1,R_2]=-\frac{i\hbar}{eb},\label{eqn:formal.QHE.NCG.guiding}
\end{equation}
see \S 8.6.2 of \cite{Thaller} or \cite{Pasquier}. As $R_1,R_2$ commute with the magnetic Laplacian $H_b$, they could be regarded as providing ``noncommuting coordinates'' for a Landau level.

Regardless of the precise justification or mathematical meaning of the \emph{operator} CCR \eqref{eqn:formal.QHE.NCG} (or \eqref{eqn:formal.QHE.NCG.guiding}), what we actually discover in QHE experiments is an \emph{exact} quantization of Hall conductance as a \emph{numerical} integer multiple of $\frac{e^2}{h}$. Correspondingly, when modelling the QHE with Landau Hamiltonians, what is rigorously proven to be exact is not a CCR commutator (as in \eqref{eqn:formal.QHE.NCG}, which varies with $b$), but rather, the numerical result of applying the trace functional to a closely related commutator. Specifically, in the Kubo Hall conductance formula, Eq.~\eqref{eqn:Kubo.formula}, the switch functions $f_1, f_2$ are normalized, bounded versions of the classical position coordinate functions $x,y$. Then it is the numerical expression \mbox{${\rm Tr}[P_{f_1},P_{f_2}]=\frac{1}{2\pi i}$}, rather than \eqref{eqn:formal.QHE.NCG}, which more accurately expresses the idea that Landau levels have ``noncommutative $x$-$y$ coordinates'', and is relevant for the quantized Hall conductance.

\medskip
More generally, the CHHP formula \eqref{eqn:CP.formula} can be interpreted as providing ``exact numerical quantization'' schemes, in the following sense. For each pair of almost-commuting bounded self-adjoint operators $A,B$, there is a representing measure $\mu_{A,B}$ such that the \emph{integrated} Poisson bracket is ``quantized'' to the \emph{traced} commutator. The adjective ``exact'' means that the correspondence is not merely at the level of formal power series, but is a true \emph{equality}.

\begin{rem}
The kilogram was recently redefined in terms of Planck's constant $h$ \cite{vonKlitzing}. The practicality of this intrinsic redefinition is contingent on having accurate and reliable access to $h$. This is \emph{not} achieved by ``measuring'' a ``microscopic'' CCR such as \eqref{eqn:CCR}. Rather, it is the exact quantization of Hall conductance as numerical multiples of $e^2/h$, together with a similarly accurate access to $\frac{2e}{h}$ via Josephson junctions, which give accurate access to $h$ \cite{vonKlitzing}; both involve \emph{macroscopic} quantum systems. Thus, on the theoretical side, it is the \emph{exact numerical quantization} of the Kubo commutator-trace (and its non-triviality for Landau levels) which is of fundamental importance for accessing $h$, whereas standard CCRs such as \eqref{eqn:CCR} or \eqref{eqn:formal.QHE.NCG.guiding} do not play a prominent role.
\end{rem}

\subsection{Fractional traces}\label{sec:rational}
Corollary \ref{cor:rational} indicates that we obtain rational commutator-traces by putting suitable polynomials $p,q$ into the formula of Theorem \ref{thm:generalization}. For example, the following higher-order versions of the chiral asymmetry formula \eqref{eqn:chiral.asymmetry} can be computed as such a commutator-trace,
\begin{align}
2\pi i\,\Tr\big((P_{f_1}P_{f_2})^{n}-(P_{f_2}P_{f_1})^{n}\big)&=2\pi i\,\Tr\big[P_{f_1}(P_{f_2}P_{f_1})^{n-1},P_{f_2}\big]\label{eqn:anyon.Kubo}\\
&=\int_S\{x^{n}y^{n-1},y\} & ({\rm Theorem}\;\ref{thm:generalization})\nonumber\\
&=\int_0^1\int_0^1 nx^{n-1}y^{n-1}\;dx\,dy \nonumber\\
&=\frac{1}{n},&n\geq 1.\label{eqn:rational.calculation}
\end{align}

We emphasize that the trace on the left side of \eqref{eqn:anyon.Kubo} has no obvious relation to $\Tr[P_{f_1},P_{f_2}]$, and it would be a formidable task to directly compute it. Even if computations succeed, say via numerics, we will be confronted with the apparent miracle of exact \emph{fractionally quantized} results. The CHHP index theory, in the form of our Theorem \ref{thm:generalization}, explains how \eqref{eqn:anyon.Kubo} is descended from $2\pi i\Tr[P_{f_1},P_{f_2}]=1$, via the principal function $G_{P_{f_1},P_{f_2}}$ and its peculiar form as a characteristic function on a unit square.

\subsubsection{Fractional conductance}\label{sec:fractional.conductance}
In this subsection, we speculate on a possible connection of the ``hidden'' fractional traces, \eqref{eqn:rational.calculation}, to the FQHE. 

Recall Eq.\ \eqref{eqn:Landau.quantization}, that one has Landau quantization of the spectrum of the Landau operator $H_b$ to the discrete set $(2\NN+1)b$. Suppose the filling factor $\nu$ equals $1$. Assuming that the electron-electron interactions are negligible, the many-body ground state when $\nu=1$ is \emph{non-degenerate}, due to the energy gap to the next-available excited state. This ground state is completely determined by the Fermi projection $P$ onto the LLL eigenspace. Informally, the electron gas is in an ``incompressible'' state. In such a situation, the Kubo formula for Hall conductance in terms of the Fermi projection $P$, Eq.\ \eqref{eqn:Kubo.formula}, is applicable, and we saw from Proposition \ref{prop:main.calculation} that the Hall conductance at filling factor $\nu=1$ is $-1/2\pi$. 

Including physical units, the Kubo formula has a factor of $\frac{e^2}{\hbar}$, where $h$ is Planck's constant and $e$ is the electron charge. It is customary to state the Hall conductance in units of $\frac{e^2}{h}$, so there is precisely one unit of Hall conductance when $\nu=1$. Similarly, each fully-occupied Landau level eigenspace contributes one unit of Hall conductance \cite{ASS,TX}. So, by additivity, at integral filling factors $\nu=\ell\in\NN$, the Hall conductance is $\ell$. 

Actually, the Hall conductance remains $\ell$ also when $\nu\approx \ell\in\NN$. This is because disorder induces localization of some of the electron states, and for the Hall conductance, it does not matter whether these localized states are filled or not. Roughly speaking, the more disorder, the more irrelevant localized states there are, so we observe broader integral plateaux in Hall conductance. So $\sigma_{\rm Hall}$ as a function of $\nu$ resembles a staircase function, with integer-valued steps centred at integer $\nu$. The Integer Quantum Hall Effect (IQHE) is fairly well-explained this way.

When the disorder is reduced, the integer-valued plateaux become narrower, and $\sigma_{\rm Hall}$ will interpolate between adjacent integer plateaux $\ell\to\ell+1$ over a wider range of filling factors $\nu$. The Fractional Quantum Hall Effect (FQHE) is the phenomenon that extra ``hidden'' Hall conductance plateaux occur at certain fractional values $\frac{c}{d}$, when $\nu\approx \frac{c}{d}$. At such non-integral $\nu$, the non-interacting electron gas model does not provide a clear spectral gap, so it might appear that the incompressible fluid picture and derivation of the Kubo formula would not apply. Why then, do plateaux in Hall conductance still appear at these non-integral $\nu$?

The resolution is that electron-electron interactions cannot be neglected when $\nu$ is not close to an integer. Generally speaking, the true many-body ground state in the presence of such interactions is impossible to determine. But for certain ``magic'' fractional filling factors, physicists have proposed very successful models for describing the many-body ground state minimizing the interaction energy, and arguing that in such situations, the ground state is actually gapped.

Let us briefly recall one such model, called the composite fermion model. As summarized by its pioneer in \cite{JainPRL}, \emph{the FQHE of electrons is the manifestation of the IQHE of composite fermions.}
Specifically, at filling factor $\nu=\frac{1}{2m+1}$ with \emph{odd} denominator, it is proposed that each electron is bound to (or ``absorbs'') $2m$ flux quanta, effectively forming weakly interacting composite fermions (``quasiparticles'') which experience the effective magnetic field strength
\[
b^*=b(1-2m\nu)=\frac{b}{2m+1}.
\]
Since the field strength is effectively suppressed by the factor $\nu=\frac{1}{2m+1}$, the density of single-quasiparticle states is likewise reduced. So the effective filling factor becomes an integer, $\nu^*=\nu/\nu=1$, and we \emph{do} have an incompressible fluid, albeit one comprising quasiparticles. For $\nu$ with larger numerators, more effective Landau levels are completely filled. Once the effective incompressibility is justified for certain filling factors $\nu=\frac{c}{d}$, the Hall conductance is expected to equal $\frac{c}{d}$, and exhibit those values as plateaux as $\nu$ is varied around $\frac{c}{d}$. Even with low disorder, plateaux formation still occurs by other pinning mechanisms \cite{Kim}.

Although the above effective description resembles the IQHE situation, it is crucial that the quasiparticle excitations are no longer standard fermions, but \emph{anyons} \cite{ASW, Halperin}, characterized by unusual exchange statistics sensitive to whether the exchange is carried out clockwise or anticlockwise on the sample surface. One even expects that the anyonic quasiparticle excitations have ``fractional charge'' $\frac{e}{2m+1}$ \cite{LaughlinNobel}. It is very significant that recent experiments have confirmed anyonic statistics for FQHE states \cite{Nakamura}. 

\medskip
For the IQHE, the (ordinary fermionic) Kubo formula has the virtue of not being a priori integer-quantized. Integrality theorems for the Kubo formula are \emph{proved}, so that there is a substantive and surprising experimental prediction that an priori real-valued measurement produces only integer-valued outcomes (whenever $\nu$ is near an integer). It is desirable to have an ``anyonic'' version of this story, i.e., an ``anyonic Kubo formula'' for which rationality theorems can be proved and applied to predict fractional Hall conductances measured in the FQHE. 

Now, at effective filling factor $\nu^*=1$, the Hall conductance is \emph{not} obtained by putting an LLL Fermi projection $P$ into the standard Kubo formula \eqref{eqn:Kubo.formula}. Indeed, doing so would lead to the wrong prediction that $\sigma_{\rm Hall}|_\nu=1$. Because of the $\frac{e^2}{h}$ factor in the Kubo formula with physical units, even if one corrects the effective charge at filling factor $\nu=\frac{1}{2m+1}$ to be $\frac{e}{2m+1}$, the wrong result $\frac{1}{(2m+1)^2}$ would be obtained.

At present, there is no established ``anyonic Kubo formula'', much less a mathematically rigorous derivation of such a formula from first principles. The expression \eqref{eqn:anyon.Kubo} is a suggestive candidate at filling factor $\nu=\frac{1}{2m+1}, m\in\NN$, since it reduces to the standard fermionic Kubo formula in the $m=0$ case, and also produces the correct fractional result when an effective Landau level projection $P$ for $\nu^*=1$ is taken as input. Furthermore, we verified that it has the highly non-obvious property of additivity with respect to the effective Landau level filling factor (Section \ref{sec:additivity}), so it naturally also handles those $\nu$ with numerators greater than 1.

\section*{Acknowledgements}
The author thanks X.~Tang, J.~Xia, and K.~Yamamoto for helpful discussions.

\section*{Statements and Declarations}
No funding was received to assist with the preparation of this manuscript. The author has no competing interests to declare that are relevant to the content of this article. 

\section*{Data availability statement}
This manuscript has no associated data.

\end{document}